\documentclass[%
 reprint,
 amsmath,amssymb,
aps,
superscriptaddress,
]{revtex4-1}
\usepackage{color}
\usepackage[utf8]{inputenc}
\usepackage{placeins}
\usepackage{graphicx}
\usepackage{dcolumn}
\usepackage{bm}
\usepackage[colorlinks=true,allcolors=blue]{hyperref}%
\usepackage[T1]{fontenc}
\usepackage[caption=false]{subfig} 

\begin{document}


\title{Electronic bulk and domain wall properties in B-site doped hexagonal ErMnO$_3$}

\author{T. S. Holstad}
 \affiliation{Department of Materials Science and Engineering, Norwegian University of Science and Technology (NTNU), NO-7491 Trondheim, Norway.}
\author{D. M. Evans}%
\email{donald.evans@ntnu.no}
\affiliation{Department of Materials Science and Engineering, Norwegian University of Science and Technology (NTNU), NO-7491 Trondheim, Norway.}%

\author{A. Ruff}
 \affiliation{Center for Electronic Correlations and Magnetism and Institute of Materials Resource Management, University of Augsburg, D-86159 Augsburg, Germany.}

\author{D. R. Småbråten}
\affiliation{Department of Materials Science and Engineering, Norwegian University of Science and Technology (NTNU), NO-7491 Trondheim, Norway.}%

\author{J. Schaab}
\affiliation{Department of Materials, ETH Zurich, 8093 Zürich, Switzerland.}%

\author{Ch. Tzschaschel}
\affiliation{Department of Materials, ETH Zurich, 8093 Zürich, Switzerland.}%

\author{Z. Yan}
\affiliation{Department of Physics, ETH Zurich, 8093 Zürich, Switzerland.}%
\affiliation{Materials Sciences Division, Lawrence Berkeley National Laboratory, Berkeley, California 94720, USA.}

\author{E. Bourret}
\affiliation{Materials Sciences Division, Lawrence Berkeley National Laboratory, Berkeley, California 94720, USA.}


\author{S. M. Selbach}
\affiliation{Department of Materials Science and Engineering, Norwegian University of Science and Technology (NTNU), NO-7491 Trondheim, Norway.}%

\author{S. Krohns}
 \affiliation{Center for Electronic Correlations and Magnetism and Institute of Materials Resource Management, University of Augsburg, D-86159 Augsburg, Germany.}

\author{D. Meier}
\affiliation{Department of Materials Science and Engineering, Norwegian University of Science and Technology (NTNU), NO-7491 Trondheim, Norway.}%



\begin{abstract}
Acceptor and donor doping is a standard for tailoring semiconductors. More recently, doping was adapted to optimize the behavior at ferroelectric domain walls. In contrast to more than a century of research on semiconductors, the impact of chemical substitutions on the local electronic response at domain walls is largely unexplored. Here, the hexagonal manganite ErMnO$_3$ is donor doped with  Ti$^{4+}$. Density functional theory calculations show that Ti$^{4+}$ goes to the B-site, replacing Mn$^{3+}$. Scanning probe microscopy measurements confirm the robustness of the ferroelectric domain template. The electronic transport at both macro- and nanoscopic length scales is characterized. The measurements demonstrate the intrinsic nature of emergent domain wall currents and point towards Poole-Frenkel conductance as the dominant transport mechanism. Aside from the new insight into the electronic properties of hexagonal manganites, B-site doping adds an additional degree of freedom for tuning the domain wall functionality.

\end{abstract}

\pacs{Valid PACS appear here}
\maketitle




\section{\label{sec:intro}Introduction}

Electronic correlation that is confined to two dimensions, found in so-called ‘2D systems’, has a large technological potential \cite{Hwang2012}. This is in part due to the electronic anisotropy, but also due to the unusual physics that has been found in these systems. Local electronic correlations are now intensively studied in a wide range of 2D materials, such as single-layer graphene \cite{Geim2007}, MoS$_2$ \cite{Desai2016}, surface states in topological insulators \cite{Zhang2012}, and oxide interfaces \cite{Hwang2012}.~A specific type of oxide interface ---  that is naturally occurring --- are domain walls (DWs) \cite{Salje2010}.~DWs show diverse confinement enabled functional properties, which are distinct from the bulk matrix: it has already been established that DWs can be magnetic \cite{ Geng_2012}, multiferroic \cite{Leo2015}, (super-) \cite{Aird1998} conductive \cite{Seidel2009, Schroder2011, Meier-2012, Wu2012, Sluka2013, Guyonnet2011, Maksymovych2012, Farokhipoor2011, McQuaid2017, Campbell2016}, and have local strain gradients (for twin walls) \cite{Carpenter2015}. These functional properties are readily influenced by electrostatics, strain, and chemical doping \cite{Gopalan2007, Catalan2012, Meier2015}. Indeed, it is the ability to control the DW behavior, combined with their sub nanometer-size \cite{Jia2008} and the ease with which they can be created and removed \cite{Whyte2014}, that has driven research interest. \newline

Since the first direct observation of conducting DWs in BiFeO$_3$ \cite{Seidel2009}, significant progress has been made on the fundamental science behind DWs. It has been established that their local properties are largely dominated by the interplay of local polarization states \cite{Eliseev2011, Gureev2011} and available charge carriers \cite{Hassanpour2016, Schaab2016-2}. Despite this progress, the research trying to produce a functional device is still in an early stage \cite{Whyte2015, Sharma2017, Mundy2017}: one of the key challenges is the optimization of emergent electronic DW behavior beyond the as-grown properties \cite{Meier2015}. Several methods of tuning the DWs have been demonstrated: FIB induced defects to control formation position \cite{Whyte2014, McGilly2017}, oxygen doping to induce ionic defects which have a propensity to form at the walls \cite{Farokhipoor2011, Gaponenko2015}, and chemical doping \cite{Hassanpour2016, Schaab2016-2}. While chemical doping with both donor and acceptor atoms is standard practice in silicon technologies, its influence on ferroelectric DWs  remains largely unexplored. \newline

An interesting model system for such doping-dependent studies is the hexagonal manganites, $R$MnO$_3$ ($R$ = Sc, Y, In, Dy to Lu). Their bulk properties are well-characterized in experiment  \cite{Fiebig2002, Lorenz2013, Meier2013} and theory \cite{vanAken2004, Fennie2005} and the system naturally provides stable charged head-to-head ($\rightarrow \leftarrow$) and tail-to-tail ($\leftarrow \rightarrow$) DWs in the as-grown state, making it an excellent model template material \cite{Meier-2012, Salje2016}. Furthermore, the system has enough chemical flexibility to allow doping, as reflected by previous investigations on the bulk level - reporting chemical substitution on the A-site \cite{Moure1999, vanAken2001-2} and B-site \cite{Ismailzade1971, Asaka2005, Levin2017}, as well as oxygen off-stoichiometry \cite{McCarthy1973, Remsen2011, Selbach2012, Skjervo2016}. Recently, the effect of doping has been extended to the micro- and nanoscopic length scales: specifically, aliovalent substitution of A-site cations was scrutinized as a control parameter for adjusting the electronic DW behavior \cite{Schaab2016-2, Hassanpour2016, Salje2016}. So far, the only spatially resolved work on B-site doping addressed high-concentration substitution in Y(Mn$_{1-x}$,Ti$_x$)O$_3$, reporting the loss of the $R$MnO$_3$-type \cite{Jungk2010} ferroelectric domain pattern for $x \gtrsim$ 17.5 \% \cite{Mori2005}. Thus, it remains an open question whether B-site doping can be used for DW engineering. \newline

\begin{figure}[ht!]
\includegraphics[scale=0.225]{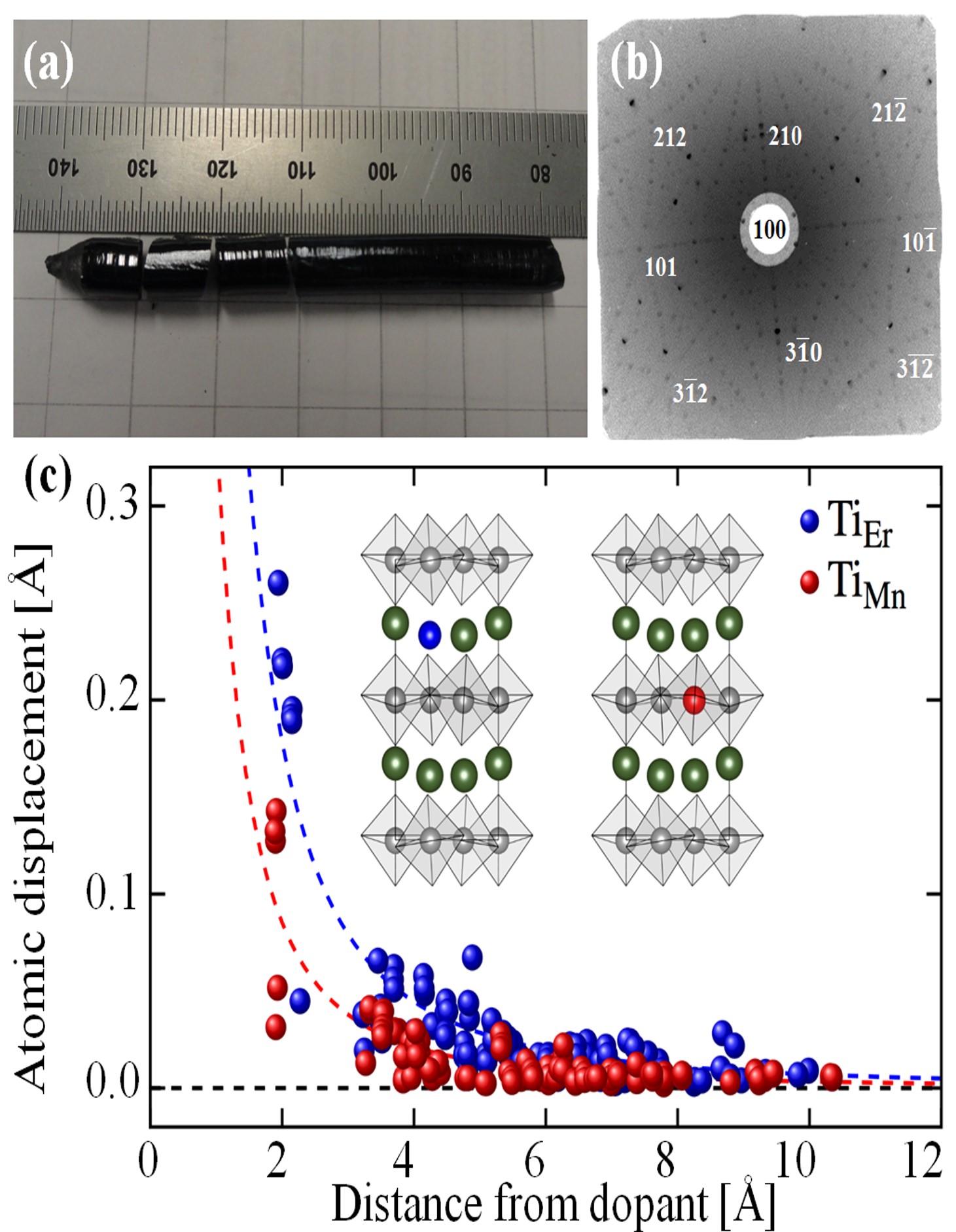}
\caption{Single-phase crystals of Er(Mn$_{1-x}$,Ti$_x$)O$_3$ ($x$ = 0.2\%). (a) Crystal ingots with a size of 5 mm in diameter and about 60 mm in length. (b) Laue back reflection data, indexed with the Cologne Laue Indexation Program (CLIP). (c) Atomic displacement dependency on whether the Ti$^{4+}$ is on the B-site (red) or A-site (blue).}
\label{Fig1}
\end{figure}
\FloatBarrier

In this work, electronic DW conductance in Ti-doped erbium manganite, Er(Mn$_{1-x}$,Ti$_x$)O$_3$ ($x$ = 0.002), is reported. By replacing Mn$^{3+}$ for Ti$^{4+}$ the bulk conductivity is reduced by an order of magnitude. The DW transport is characterized using I(V)-spectroscopy and time-dependent measurements, confirming the intrinsic nature of the DW currents. Temperature-dependent I(V) measurements support Poole-Frenkel conduction as the predominant conduction mechanism. This work expands the chemical parameter space for DW property engineering in $R$MnO$_3$ by establishing Ti$^{4+}$ as a B-site donor dopant.


\FloatBarrier
\section{\label{sec:exp}Experimental details}

The p-type \cite{SubbaRao1971,Skjervo2016} semiconductor ErMnO$_3$ is used as the parent material in this work. It has a hexagonal $P6_3cm$ crystal structure and displays improper ferroelectricity  at room-temperature  (T$_c \approx $ 1150 $^o$C)  \cite{vanAken2004, Chae2012}. The spontaneous polarization in  $R$MnO$_3$ is  P $\approx$ 5.5 $\mu$C/cm$^{2}$ , pointing along the c-axis \cite{Smolenskii1964, vanAken2004}. High-quality single-phase crystals of the compound, hexagonal  Er(Mn$_{1-x}$,Ti$_x$)O$_3$ ($x$ = 0.002), are grown by the pressurized floating-zone method \cite{Yan2015} (see Fig.~\ref{Fig1} (a)). After confirming the anticipated hexagonal target phase (not shown), the Er(Mn$_{1-x}$,Ti$_x$)O$_3$ crystal is oriented by Laue diffraction and cut into disc-shaped platelets with a thickness of $\sim$1 mm and an in-plane polarization. Representative Laue back reflection data is shown in Fig.~\ref{Fig1} (b), confirming the single-crystallinity of the sample. After preparing oriented samples with in-plane polarization, chemo-mechanical polishing with silica slurry is applied, which yields flat surfaces with a root mean square roughness of about 0.5-1.5 nm and improves the quality of the subsequent analysis by scanning probe microscopy (SPM). \newline

The SPM measurements are performed with a NT-MDT Ntegera Prima SPM.  Piezo-response force microscopy (PFM) data are collected  at room temperature using Stanford Research 830R lock-in amplifiers and applying an AC voltage to the tip ($\omega$ = 40 kHz, $U_{RMS}$ = 5 V).  Conducting atomic force microscopy (cAFM) and local IV-spectroscopy measurements are performed by applying a positive bias to the sample while grounding the tip. PFM, cAFM and IV-spectroscopy measurements were performed using a $\mu$masch NSC35:HQ hard diamond-like carbon coated tip. Electrostatic force microscopy (EFM) images are collected at room temperature using Stanford Research 830R lock-in amplifiers and a NT-MDT DCP20 tip with nitrogen doped diamond coating. The EFM data is recorded at a frequency $\omega$, while scanning the sample in non-contact mode with an AC voltage applied to the tip ($\omega$ = 18.7 kHz, $U_{pp}$ = 20 V) and the sample grounded. The dielectric properties at frequencies from 1 Hz $\leq$ $\nu$ $\leq$ 1 MHz are determined using a frequency-response analyzer (Novocontrol AlphaAnalyzer). For these dielectric analyses, contacts of silver paste and wires in a pseudo-four-point configuration are applied to opposite faces of the plate-like samples. The measurements are performed between 50 K and 300 K in a closed-cycle refrigerator with the samples in vacuum.

\section{\label{sec:dft}Density Functional Theory Calculations}
While the structural characterization confirms the anticipated hexagonal target phase of the moderately doped  Er(Mn$_{1-x}$,Ti$_x$)O$_3$ sample (Fig. \ref{Fig1} (a)), it cannot decide whether Ti$^{4+}$ occupies the Mn- or Er-sublattice.  Density functional theory (DFT) calculations are therefore performed to investigate which cation sublattice is preferred for Ti$^{4+}$. The DFT calculations are performed with the projector augmented wave method (PAW) \cite{Blochl1994}, as implemented in VASP \cite{Kresse1996, Kresse1999}, using the PBEsol exchange correlation functional \cite{Perdew2008} to study the local structural changes and energetics upon Ti-doping of ErMnO$_3$ in both cation sublattices. Er 5p, 4f, 5d and 6s (with 11 f-electrons frozen in the core), Mn 3p, 4s and 3d, O 2s and 2p, and Ti 3s, 3p, 3d and 4s are treated as valence electrons. 2x2x1 supercells with one Ti ion per supercell, (Er$_{1-x}$,Ti$_x$)MnO$_3$ or Er(Mn$_{1-x}$,Ti$_x$)O$_3$ ($x=1/24$), are chosen as the model systems. Plane wave energy cutoff is set to 550 eV, and the Brillouin zone is sampled with a $\Gamma$-centered 2x2x2 k-point grid. GGA+U \cite{Dudarev1998} with U of 5 eV applied to Mn 3d is used to reproduce the experimental band gap. The Mn sublattice is initialized with collinear frustrated antiferromagnetic (f-FAFM) order \cite{Medvedeva2000}. The lattice positions are relaxed until forces on all atoms are below 0.005 eV/Å. The defect formation energy is calculated by $E_{\textrm{def}}^f=E_{\textrm{defect}} -E_{\textrm{ref}} -\sum_i\left[ n_i \cdot \mu_i \right]$, where $E_{\textrm{defect}}$ and E$_{\textrm{ref}}$ are the energies of the defect cell and stoichiometric ErMnO$_3$, respectively (n$_i$ is number of species $i$ added per supercell, and $\mu_i$ the chemical potential of species $i$). The chemical potentials of Er, Mn, and Ti are defined by the chemical equilibria between the binary oxides Er$_2$O$_3$, Mn$_2$O$_3$ and TiO$_4$ through Er$_2$O$_3\rightleftharpoons$ 2Er+3O, Mn$_2$O$_3\rightleftharpoons$ 2Mn+3O and TiO$_2\rightleftharpoons$ Ti+2O. Whence, E$_{\textrm{def}}^f$ becomes a function of the oxygen chemical potential, i.e., the oxygen partial pressure during synthesis.\newline

Intuitively, it is likely that Ti$^{4+}$ replaces Mn$^{3+}$ due to the similar atomic radii ($r_{\textrm{Er}^{3+}\textrm{(VII)}} = $ 0.945 Å, $r_{\textrm{Mn}^{3+}\textrm{(V)}} = $ 0.58 Å, and $r_{\textrm{Ti}^{4+}\textrm{(V)}} = $ 0.51 Å according to Shannon \cite{Shannon1976}). The structural distortion profiles upon doping is plotted as a function of distance from the dopant in Fig.~\ref{Fig1} (c). As a guide for the eye, the profiles are fitted to $\propto r^{–2}$, marked in dashed lines. The calculations show that local displacements close to the dopant are smaller for B-site substitution compared to A-site substitution, as expected from the cation size mismatches. Even though the differences are quite subtle, on the order of 0.1 Å closest to the dopant, the decreased structural distortions by B-site substitution demonstrate a site preference of Ti$^{4+}$ for the B-site. This is further confirmed by the calculated defect formation energies, where the defect formation energy of B-site substitution is 0.79 eV lower than that of A-site substitution. It is to be noted that the calculated defect formation energies strongly depend on the definition of the cation chemical potentials. However, with the binary oxides Er$_2$O$_3$ and Mn$_2$O$_3$ constituting ErMnO$_3$, and TiO$_2$ as choice of definition, it still gives a good qualitative trend for the B-site preference with respect to defect formation energetics.

\FloatBarrier
\section{\label{sec:dielectric}Dielectric spectroscopy}
\begin{figure}[ht!]
\includegraphics[scale=0.21]{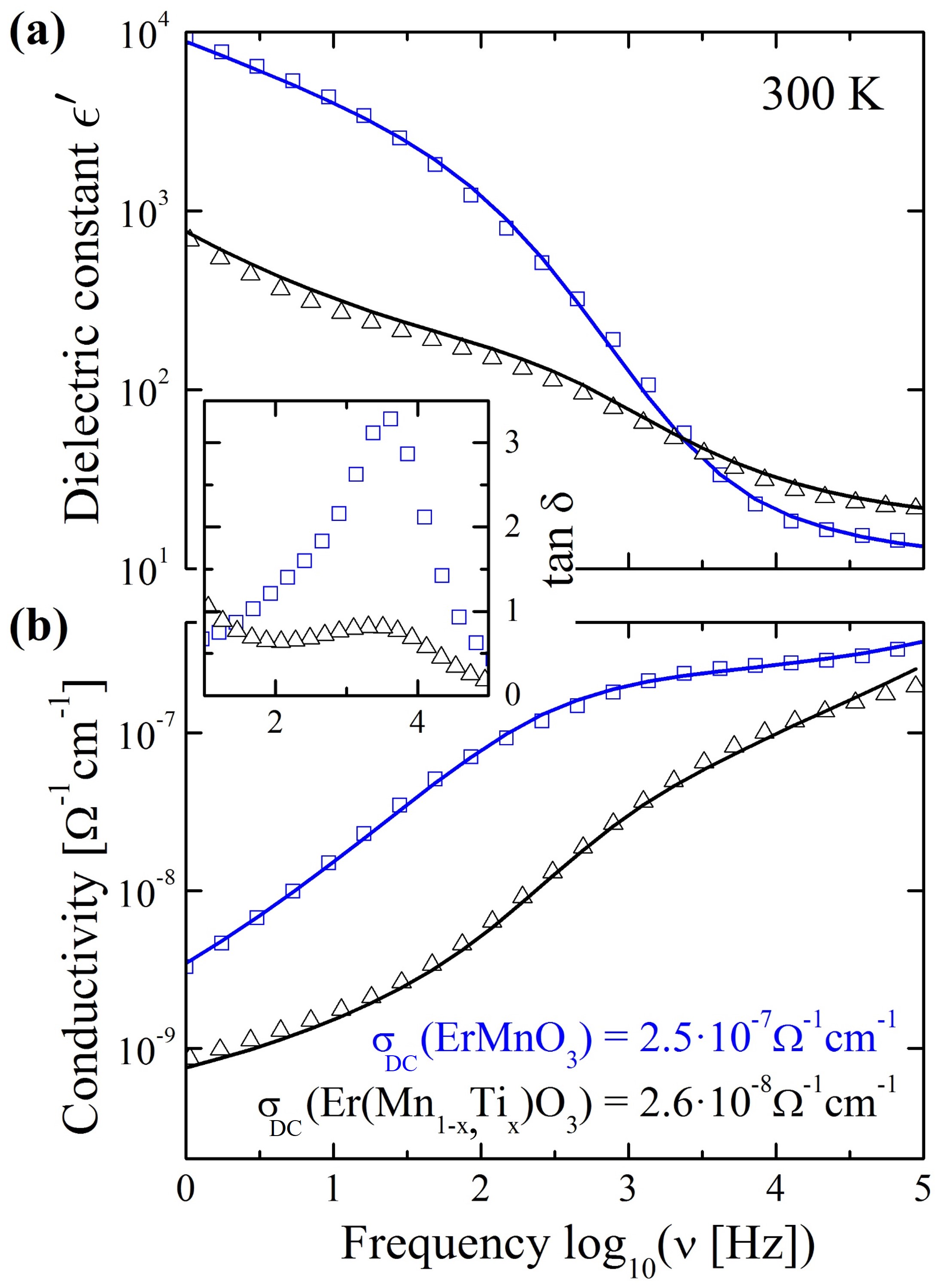}
\caption{Dielectric response of ErMnO$_3$ and Er(Mn$_{1-x}$,Ti$_x$)O$_3$ ($x$ = 0.2\%), blue and black data points, respectively. (a) Dielectric constant as a function of frequency. Tan($\delta$) data as an inset in the same frequency range. (b) Frequency dependent conductivity. Lines are fits from equivalent circuits. }
\label{Fig2}

\includegraphics[scale=0.16]{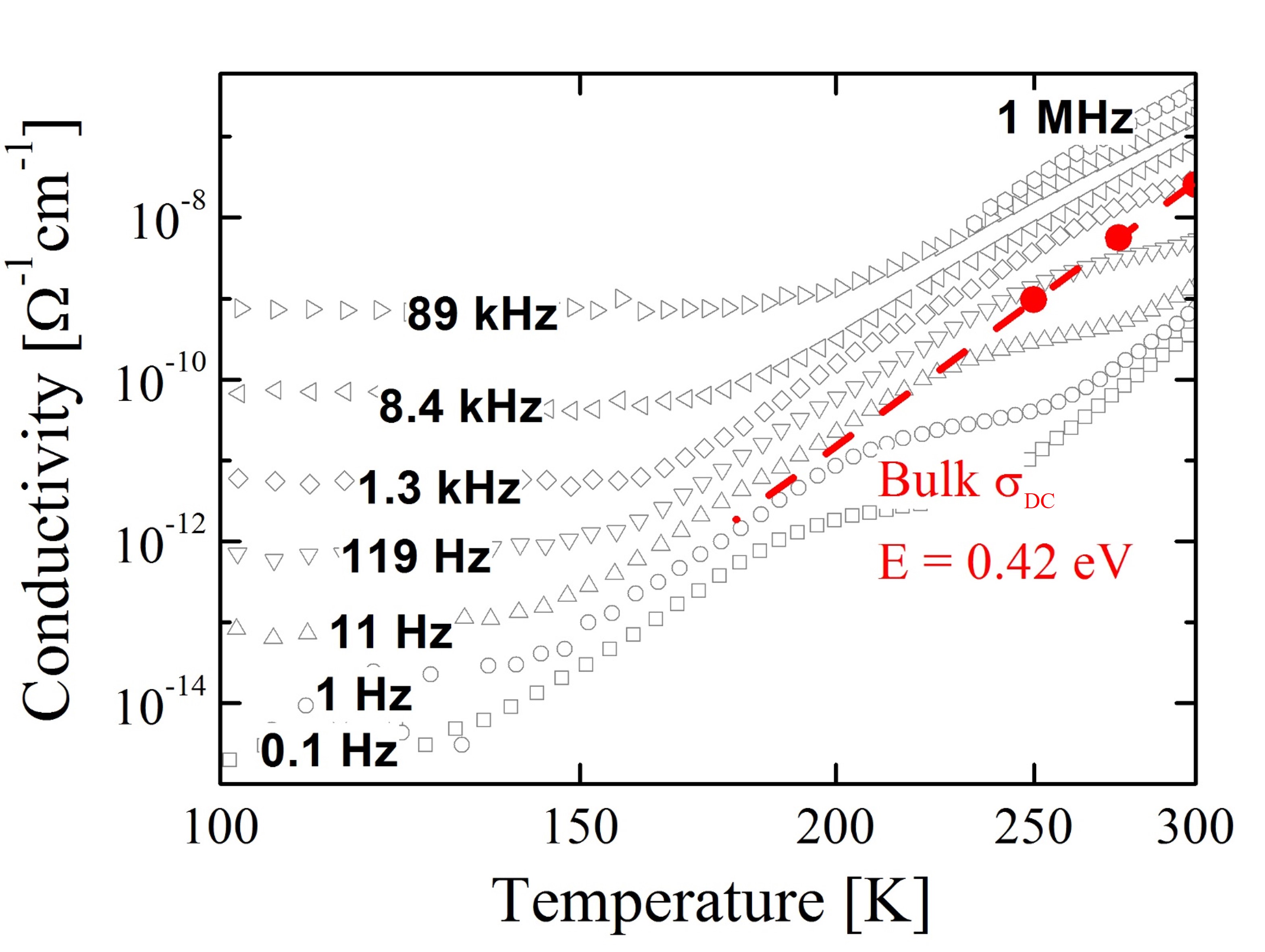}
\caption{Conductivity of single-phase Er(Mn$_{1-x}$,Ti$_x$)O$_3$ ($x$ = 0.2\%) in an Arrhenius representation for various frequencies. The closed symbols denote the DC conductivity revealed by equivalent circuit analysis of the frequency dependent spectra. }
\label{Fig3}
\end{figure}

After corroborating the replacement of Mn$^{3+}$ by Ti$^{4+}$, the impact of B-site doping on the bulk electronic properties is investigated. In order to gain quantitative information, dielectric spectroscopy measurements are performed on ErMnO$_3$ and Er(Mn$_{1-x}$,Ti$_x$)O$_3$; Fig.~\ref{Fig2} blue and black data points, respectively. The dielectric constant for both samples shows a stepwise decrease (from ca. 10$^4$-10$^3$ at 1 Hz, to about 10-20 at 100 kHz, respectively) and associated peaks in tan($\delta$) (see inset in Fig.~\ref{Fig2}). This behavior is typical of a relaxation process. The relaxation is likely from an electrical heterogeneous phase, often termed Maxwell-Wagner relaxation, and possible mechanisms include: surface barrier layers formed by Schottky diodes \cite{Krohns2008} or internal barrier layers \cite{Lunkenheimer2010}, e.g. DWs. The tan($\delta$) for the relaxation process is 3 for the undoped crystal and 1 for Ti-doped, both typical values for Maxwell-Wagner relaxations where tan($\delta$)  $\gtrsim$ 1 is to be expected.

\FloatBarrier

The data is fitted with a simple equivalent circuit model (consisting of two RC-elements in series: one for the bulk; and one for a layer capacitance) for the undoped sample; the fit is given by the blue solid line in Fig.~\ref{Fig2} (a). The Ti-doped sample needs an additional RC-element because of a low frequency ($\sim$1 Hz) relaxation, given by the black solid line. The origin of this additional relaxation, which is the subject of further work, is likely associated with DW mobility, or the Ti dopant. These fits give intrinsic dielectric constant values of $\epsilon'$ $\sim$ 11 for ErMnO$_3$ and  of $\epsilon'$ $\sim$ 22 for Er(Mn$_{1-x}$,Ti$_x$)O$_3$, which are comparable to the values for YMnO$_3$ \cite{Smolenskii1964, Adem2015}. \newline

The frequency dependent conductivity, $\sigma'$, of both samples,  Fig.~\ref{Fig2} (b), is used to evaluate doping dependent changes in the DC conductivity. The DC conductivity is given by a plateau in the frequency dependent conductivity when the RC-elements of the interface layer are short-circuited by the frequency of the applied electric field. Increases of $\sigma'$ at higher frequencies can arise from hopping charge transport. This is modeled in the equivalent circuit with an additional frequency dependent resistive element in parallel to the RC-circuit for the bulk, i.e. $\sigma' \propto \nu^s$, with $s < 1$, and corresponds to Jonscher’s universal dielectric response \cite{Jonscher1977}. These fits give intrinsic DC conductivity values of $\sigma_{DC} = 2.5 \cdot 10^{-7}$ $\Omega^{-1}$ cm$^{-1}$ and $2.6 \cdot 10^{-8}$ $\Omega^{-1}$ cm$^{-1}$, for the undoped and doped samples, respectively. The observed behavior is consistent with decreasing conductivity  in p-type h-$R$MnO$_3$ with electron doping. \newline

After establishing the differences in conductivity between the parent material and the doped crystals, and measuring values for the dielectric constant, the low temperature frequency dependency of the conductivity of the Ti-doped sample is presented in Fig.~\ref{Fig3}.~The conductivity decreases from 300 K before flattening out at ca.~150 K. 

\FloatBarrier
\onecolumngrid
\begin{center}
\begin{figure}[ht!]
\includegraphics[scale=0.19]{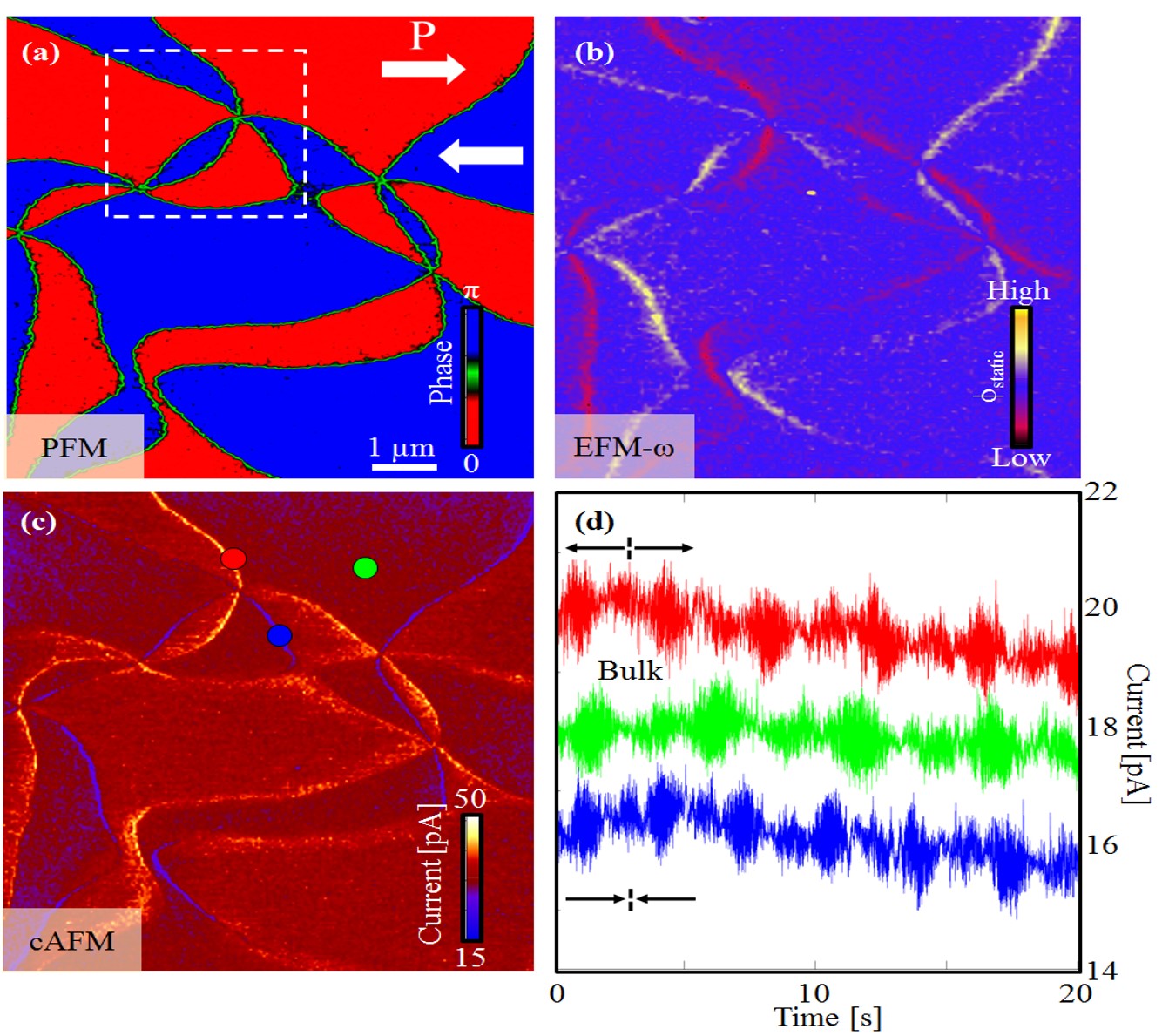}
\caption{Spatially resolved electronic properties. (a) PFM image showing the orientation of the ferroelectric polarization, indicated by white arrows. The area marked with white dashed lines is the area used in Fig.~\ref{Fig5}. (b) EFM $\omega$-channel showing a reduction (red) of the electrostatic potential  $\phi$ at the tail-to-tail DWs and an enhancement (yellow) at the head-to-head DWs. (c) cAFM image collected at 30 V showing enhanced conductivity at tail-to-tail DWs (yellow) and suppressed conductivity and head-to-head DWs (blue). (d) Time-dependent conductivity measurements showing stable signals for at least 20 seconds at  $U = 24 $ V. }
\label{Fig4}
\end{figure}
\end{center}
\twocolumngrid
\FloatBarrier

Furthermore, it is strongly frequency dependent at 100 K, varying by about $\sim$6 orders of magnitude across a frequency range of 0.1 Hz - 89 kHz. The red symbols indicate the intrinsic bulk DC conductivity revealed from fits with an equivalent circuit \cite{Lunkenheimer2010}, and the temperature dependency is approximated by the dashed red line. \newline

The observed behavior is further evidence for a temperature dependent activation mechanism \cite{Ruff2017, Moure1999-2}. The gradient of the DC conductivity is used to calculate an activation energy for the conduction process of E$_A$ = 0.42 eV, which is in excellent agreement with values in the literature for YMnO$_3$ of order 0.36 eV. These macroscopic measurements show that the addition of relatively small amounts of Ti$^{4+}$ to ErMnO$_3$ can have profound effects on the bulk conductivity. Next, the effects of Ti on the local conductivity properties of the bulk and DWs are investigated.



\section{\label{sec:SPM}Scanning Probe Microscopy}
The domain and DW structures in  Er(Mn$_{1-x}$,Ti$_x$)O$_3$ are investigated using SPM. Previous investigations on Y(Mn$_{1-x}$,Ti$_x$)O$_3$ ($x = $ 0.05 to 0.4) revealed that the domain structure strongly depends on the Ti-content, and B-site doping \cite{Mori2005} was found to completely destroy the $R$MnO$_3$-characteristic ferroelectric domain structure for $x \gtrsim 0.175$. Fig.~\ref{Fig4} (a) shows a representative PFM image of  Er(Mn$_{1-x}$,Ti$_x$)O$_3$ ($x = 0.002$). The in-plane contrast with red and blue areas corresponds to ferroelectric domains of opposite polarization direction with P pointing to the right and left, respectively. It is clear that the parent compound and the Ti-doped sample exhibit qualitatively equivalent domain patterns. The electrostatics at the DWs in  Er(Mn$_{1-x}$,Ti$_x$)O$_3$ are analyzed in Fig.~\ref{Fig4} (b), showing an EFM scan obtained at the same position as the PFM image in Fig.~\ref{Fig4} (a). The EFM-$\omega$ map in Fig.~\ref{Fig4} (b) reveals the distribution of bound carriers as explained in Refs.~\cite{Schaab2016-2,Johann2010-2}. A reduction (red) of the electrostatic potential  $\phi$ at the tail-to-tail DWs and an enhancement (yellow) at head-to-head  DWs is observed. That is: $ \phi_{\leftarrow \rightarrow} < \phi_{\textrm{bulk}} < \phi_{\rightarrow \leftarrow}$. Figures \ref{Fig4} (a-c) thus demonstrates that the distribution and electrostatics of the DWs are unaffected by the applied B-site doping, proving its usability for \mbox{electronic adjustments.} \newline

\begin{figure}[ht!]
 \includegraphics[scale=0.12]{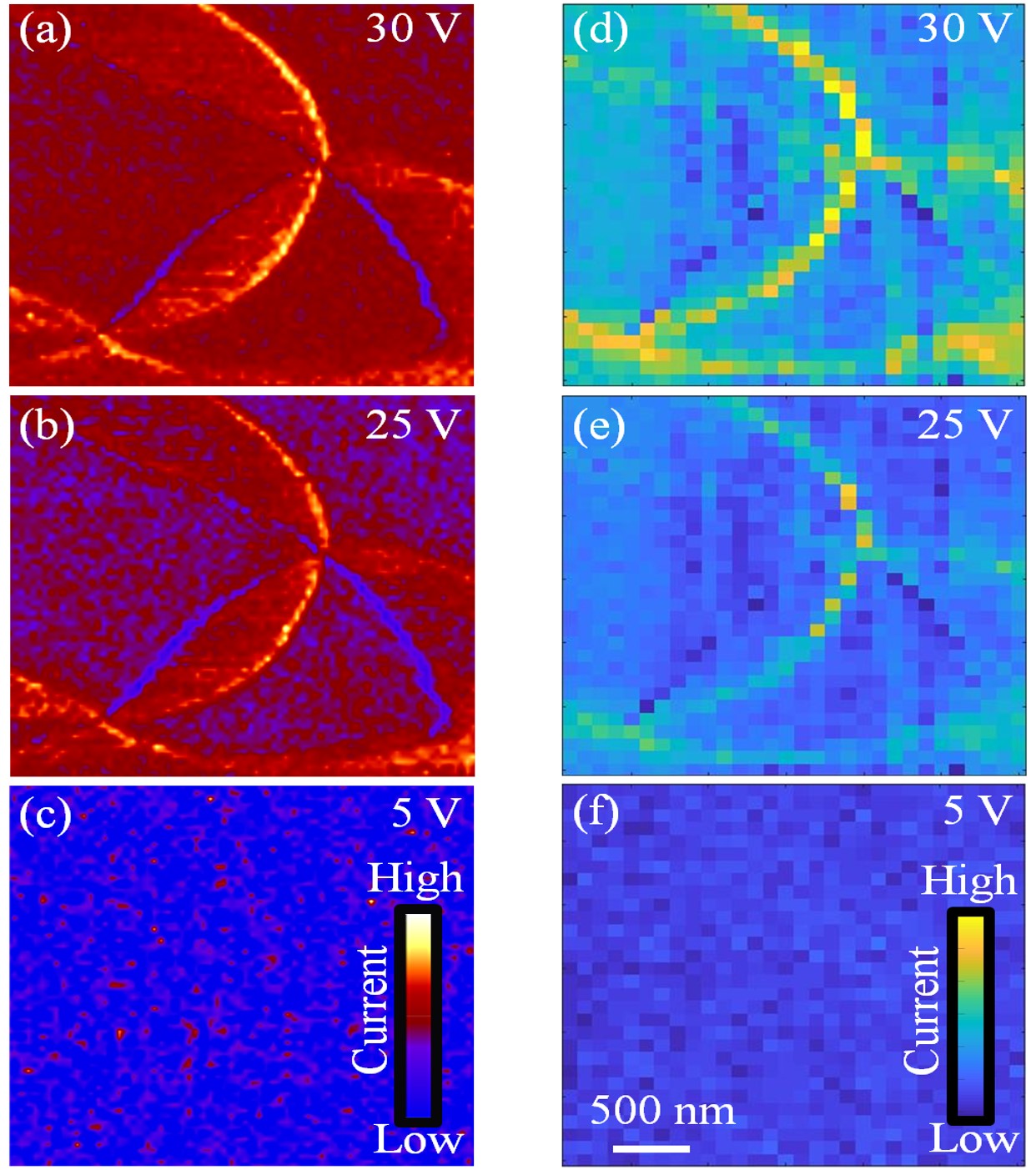}
\caption{(a), (b), and (c): c-AFM scans collected at 30 V, 25 V  and 5 V, respectively. (d), (e), and (f): IV-spectroscopy grids taken at the same position as in (a), (b), and (c) were used to construct current maps at 30 V, 25 V, and 5 V, respectively. Note that this data was taken from the area marked with white dashed lines in Fig.~\ref{Fig4} (a).}
\label{Fig5}
\end{figure}

The effect of the Ti-doping on the conduction properties of the DWs is shown in Fig.~\ref{Fig4} (c).  Qualitatively, the DWs exhibit the previously observed behavior \cite{Meier-2012, Wu2012}: tail-to-tail DWs show an enhanced conductance, while head-to-head DWs show diminished conductance with respect to the bulk.  Time-dependent current measurements, gained with a stationary tip, show a minor decrease over time  (Fig.~\ref{Fig4} (d)), indicating that  the DW conductance is a predominately intrinsic phenomenon. \newline

 To further exclude contributions from transient currents or DW movements, cAFM mapping and IV-spectroscopy measurements \cite{Maksymovych2012} are compared in Fig \ref{Fig5}. In Fig.~\ref{Fig5} (a-c), standard voltage-dependent cAFM scans are shown, recorded while scanning the tip at selected voltages. The scans are from the boxed region in Figs.~\ref{Fig4} (a). Coarser conductance maps reconstructed from I(V) spectroscopy measurements are displayed in Figs.~\ref{Fig5} (d-f). In the latter case, the tip moves to different points in a pre-defined grid and is then stationary while measuring I(V) curves at each point (this was done in a 3$\mu$m x 3$\mu$m box with 100 nm between each IV-curve and a voltage ramp rate of 5 V/sec). Independently of the applied SPM method, qualitatively equivalent results are observed, i.e. conducting tail-to-tail DWs and insulating head-to-head DWs. This  qualitative agreement corroborates that the obtained DW currents are intrinsic. \newline

\begin{figure}[ht!]
\includegraphics[scale=0.14]{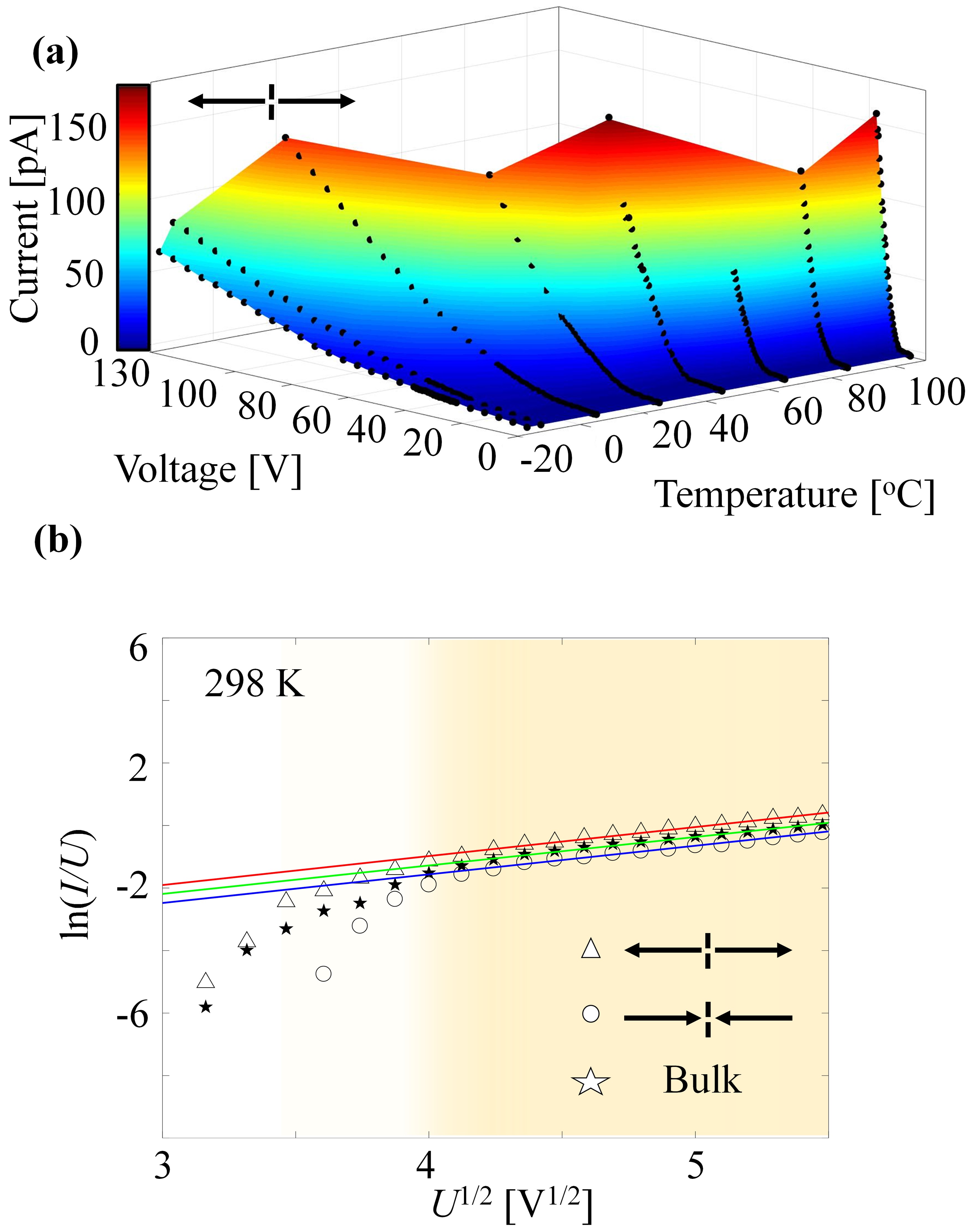}
\caption{Temperature dependency. (a) cAFM scans at different temperatures and voltages where preformed and an average value for the DW was extracted for each scan in order to produce the surface plot of I(U, T) at the tail-to-tail DWs. The black dots are the data points and the colored surface is created by interpolation between the points. (b) The PFC mechanism (\ref{eq:PFC}) gives good fits to the cAFM data in Fig.~\ref{Fig5}, but anomalous values of the dielectric constant of $\sim$ 650 at 25 $^o$C.}
\label{Fig6}
\end{figure}

Now that the intrinsic nature of the DW conduction is established, the mechanism for that conduction is investigated using temperature dependent measurements (Fig \ref{Fig6} (a)). Previous studies reported in Ref.~\cite{Mundy2017} already narrowed down the possible conduction mechanisms to Space Charge Limited Conduction (SCLC) and Poole-Frenkel Conduction (PFC). Thus, these two mechanisms are investigated in the following.~The current-voltage data in Fig.~\ref{Fig6} (a) (black dots) is extracted from cAFM scans.~For this purpose, a series of voltage-dependent images were taken at different temperatures.~The evaluation of the data shows that the bulk and the DWs behave in qualitatively the same way. As such, only the conducting tail-to-tail DWs are considered in Fig.~\ref{Fig6} (a). \newline

The most striking feature of the data is the dramatic increase in conductivity with increasing temperature. This increase is in direct contrast to SCLC, which goes as $I \propto V^{ \left( \frac{T^*}{T} + 1\right)}$, thus excluding it as a possible conduction mechanism. $T^*$ is the characteristic temperature describing the distribution of trap states in the band gap, as explained in Ref.~\cite{Simmons1971}. The obtained temperature dependency is, however, consistent with PFC: the dominant conduction mechanism in YMnO$_3$~\cite{Moure1999-2}. The latter is given by \cite{Simmons1971}, 

\begin{equation}
\label{eq:PFC}
I \propto E e^{-q \frac{\Phi_T - \sqrt{\frac{qE}{\pi \epsilon_0 \epsilon_r}}}{k_BT}}, 
\end{equation}
where $q$ is the electronic charge, $\Phi_T$ is the trap energy level, $\epsilon_0$ is the vacuum permittivity, and the electric field is estimated by $E \approx \frac{V}{r_{\textrm{tip}}}$ (Ref.~\cite{Stolichnov2015}).  The PFC model is in qualitative agreement with the room temperature data for voltages $U^{1/2} \gtrsim$ 4 V$^{1/2}$ (Fig.~\ref{Fig6} (b)), but it consistently overestimates the dielectric constant. At room temperature, PFC fits yields a dielectric constant of $\sim 650$, whereas the bulk value is measured to be $\sim 22$  (see Fig.~\ref{Fig2}). Thus, while the cAFM data in Fig.~\ref{Fig6} allows PFC to be identified as the predominant conduction mechanism in  Er(Mn$_{1-x}$,Ti$_x$)O$_3$, it is not suitable to reliably determine associated material parameters.


\section{Conclusion}

In conclusion, B-site doping has been established as a new control parameter for engineering the electronic properties at ferroelectric domain walls in $R$MnO$_3$. As a model case, Ti-doped ErMnO$_3$ was considered. DFT confirmed that Ti occupies the B-site, where it acts as a donor, reducing the bulk conductivity as quantified by dielectric spectroscopy. cAFM demonstrated the intrinsic nature of domain-wall currents, and pointed towards Poole-Frenkel as the dominant conduction mechanism. The possibility of B-site doping, in addition to previously reported A-site substitution \cite{Schaab2016-2}, enhances the electronic flexibility of the hexagonal manganites and expands the chemical parameter space available for adjusting and optimizing the electronic domain wall behavior.

\section{Acknowledgements}
The authors thank Trygve Magnus Ræder for his assistance in automating data extraction from the temperature dependent measurements. T.S.H., D.M.E., and D.M. acknowledge funding through NTNU's Onsager Fellowship Program and Outstanding Academic Fellows Programme. D.R.S. acknowledges funding through The Research Council of Norway (FRINATEK project no.~231430/F20) and NTNU. Computational resources were provided by UNINETT Sigma2 - the National Infrastructure for High Performance Computing and Data Storage in Norway (projects NTNU243 and NN9264K). Crystals were grown at the Lawrence Berkeley Laboratory supported by the U.S. Department of Energy, Office of Science, Basic Energy Sciences, Materials Sciences and Engineering Division (Contract No.~DE-AC02-05-CH11231). J.S.~and C.T.~acknowledge funding from ETH Zurich and SNF (proposal no.~200021\_149192, J.S.; proposal no.~200021\_147080, C.T.). C.T. acknowledges support by FAST, a division of the SNF NCCR MUST.

\FloatBarrier
\newpage
\bibliography{refs}{}
\bibliographystyle{apsrev4-1}

\end{document}